\documentclass[12pt,preprint]{aastex}

\shorttitle{A MHD Boost for Relativistic Jets}
\shortauthors{Mizuno et al.}

\usepackage{graphicx}

\begin{document}

\title{A Magnetohydrodynamic Boost for Relativistic Jets}

\author{
Yosuke Mizuno\altaffilmark{1,6}, Philip Hardee\altaffilmark{2},
Dieter H. Hartmann\altaffilmark{3}, Ken-Ichi Nishikawa\altaffilmark{1,4},
and \\Bing Zhang\altaffilmark{5} }

\altaffiltext{1}{National Space Science and Technology Center, 320
Sparkman Drive, VP 62, Huntsville, AL 35805, USA;
Yosuke.Mizuno@msfc.nasa.gov} \altaffiltext{2}{Department of Physics
and Astronomy, The University of Alabama, Tuscaloosa, AL 35487, USA}
\altaffiltext{3}{302C Kinard Laboratory, Clemson University,
Clemson, SC 29634-0978, USA} \altaffiltext{4}{Center for Space
Plasma and Aeronomic Research, University of Alabama in Huntsville,
Huntsville, AL 35899, USA}
\altaffiltext{5}{Department of Physics, University of Nevada, Las
Vegas, NV 89154, USA} \altaffiltext{6}{NASA Postdoctoral Program
Fellow/ NASA Marshall Space Flight Center}

\shorttitle{A MHD Boost for Relativistic Jets}
%\shortauthor{MIZUNO, ET AL.}

\begin{abstract}

We performed relativistic magnetohydrodynamic simulations of the
hydrodynamic boosting mechanism for relativistic jets explored by
Aloy \& Rezzolla (2006) using the RAISHIN code. Simulation results
show that the presence of a magnetic field changes the properties
of the shock interface between the tenuous, overpressured jet ($V^{z}_{j}$)
flowing tangentially to a dense external medium.
Magnetic fields can lead to more efficient acceleration of the jet, in
comparison to the pure-hydrodynamic case. A ``poloidal'' magnetic field
($B^{z}$), tangent to the interface and parallel to the jet flow,
produces both a stronger outward moving shock and a stronger inward moving
rarefaction wave. This leads to a large velocity component normal to
the interface in addition to acceleration tangent to the interface, and
the jet is thus accelerated to larger Lorentz factors than those obtained
in the pure-hydrodynamic case. Likewise, a strong ``toroidal'' magnetic
field ($B^{y}$), tangent to the interface but perpendicular to the jet
flow, also leads to stronger acceleration tangent to the shock interface
relative to the pure-hydrodynamic case. Overall, the acceleration
efficiency in the ``poloidal'' case is less than that of the 
``toroidal''
case but both geometries still result in higher Lorentz factors than the
pure-hydrodynamic case. Thus, the presence and relative orientation of a
magnetic field in relativistic jets can significant modify
the hydrodynamic boost mechanism studied by Aloy \& Rezzolla (2006).

\end{abstract}

\section{Introduction}

Relativistic jets have been observed in active galactic nuclei (AGN)
and quasars (e.g., Urry \& Pavovani 1995; Ferrari 1998), in black hole
binaries (microquasars) (e.g., Mirabel \& Rodiriguez 1999), and are
also thought to be responsible for the jetted emission from gamma-ray
bursts (GRBs)(e.g., Zhang \& M\'{e}sz\'{a}ros 2004; Piran 2005;
M\'{e}sz\'{a}ros 2006). Proper motions observed in jets from microquasars
and AGNs imply jet speeds from $\sim 0.9 c$ up to $\sim 0.999 c$, and
Lorentz factors in excess of $\Gamma \sim$ 100 have been inferred for GRBs.
The acceleration mechanism(s) capable of boosting jets to such
highly-relativistic speeds has not yet been fully established.

The most promising mechanisms for producing relativistic jets
involve magnetohydrodynamic centrifugal acceleration and/or magnetic
pressure driven acceleration from the accretion disk around compact
objects (e.g., Blandford \& Payne 1982; Fukue 1990), or direct
extraction of rotational energy from a rotating black hole (e.g.,
Penrose 1969; Blandford \& Znajek 1977). Recent General Relativistic
Magneto-Hydrodynamic (GRMHD) simulations of jet formation in the vicinity
of strong gravitational field sources, such as black holes or neutron
stars, show that jets can be produced and accelerated by the presence
of magnetic fields which are significantly amplified by the rotation
of the accretion disk and/or the frame-dragging of a rotating black hole
(e.g., Koide et al. 1999, 2000; Nishikawa et al. 2005; Mizuno et al. 2006b;
De Villiers et al. 2005; Hawley \& Krolik 2006; McKinney \& Gammie 2004;
McKinney 2006). The presence of strong magnetic fields is likely in areas
close to the formation and acceleration region of relativistic jets.
In the context of GRBs, standard scenarios invoke a
fireball that is accelerated by thermal pressure during the initial free
expansion phase (e.g., M\'esz\'aros et al. 1993; Piran et al. 1993).
Magnetic dissipation may occur during the expansion, and a fraction
of the dissipated energy may be used to further accelerate the fireball
(e.g., Drenkhahn \& Spruit 2002). Whether one considers the fate of the
collapsing core of a very massive star (Woosley 1993), or the merger of
a neutron star binary system (Paczynski 1986), the differentially rotating
disks that feed the newly born black hole are likely to amplify any present
seed field through magnetic braking and the magnetorotational instability
(MRI) proposed by Balbus $\&$ Hawley (1991, 1998). Numerical solutions of
the coupled Einstein-Maxwell-MHD equations (e.g., Stephens et al. 2006, and
references therein) confirm the expected growth of seed fields, even to the
point at which the fields become strong enough to be dynamically important.

Recently, Aloy \& Rezzolla (2006) investigated a potentially powerful
acceleration mechanism in the context of purely hydrodynamical flows,
posing a simple Riemann problem.  If the jet is hotter and at much
higher pressure than a denser colder external medium, and moves with a
large velocity tangent to the interface, the relative motion of the
two fluids produces a hydrodynamical structure in the direction
perpendicular to the flow (normal to the interface), composed of a
``forward shock'' moving away from the jet axis, and a ``reverse
shock'' (or a rarefaction wave) moving toward the jet axis. Aloy \&
Rezzolla (2006) label this pattern either
$_{\leftarrow}SCS_{\rightarrow}$, or
$_{\leftarrow}RCS_{\rightarrow}$), where $_{\leftarrow}S$ refers to
the reverse shock, ($_{\leftarrow}R$ to the reverse rarefaction wave),
$S_{\rightarrow}$ to the forward shock, and $C$ to the contact
discontinuity between the two fluids. In the case
$_{\leftarrow}RCS_{\rightarrow}$, the rarefaction wave propagates into
the jet and the low pressure wave leads to strong acceleration of the
jet fluid into the ultrarelativistic regime in a narrow region near
the contact discontinuity. This hydrodynamical boosting mechanism is
very simple and powerful, but is likely to be modified by the effects
of magnetic fields present in the initial flow, or generated within
the shocked outflow.

Here, we investigate the effect of magnetic fields on the boost mechanism
proposed by Aloy \& Rezzolla (2006). We find that the presence of magnetic
fields in the jet can provide even more efficient acceleration of the jet
than possible in the pure-hydrodynamic case. The highly significant role
magnetic fields may play in accretion flows (e.g., Miller et al. 2006) and
in core-collapse supernovae (e.g., Woosley $\&$ Janka 2005) is perhaps echoed
in the collimated relativistic outflows from some compact stellar remnants.

\section{Numerical Method}

In order to study the magnetohydrodynamic boost mechanism for
relativistic jets, we use a 1-dimensional special relativistic MHD
(RMHD) version of the 3-dimensional GRMHD code ``RAISHIN'' in
Cartesian coordinates (Mizuno et al. 2006a). A detailed description of
the code and its verification can be found in Mizuno et al. (2006a).
In the simulations presented here we use the piecewise parabolic
method for reconstruction, the Harten, Lax, \& van Leer (HLL)
approximate Riemann solver (Harten et al. 1983), a flux constrained
transport scheme to keep the magnetiic fields divergence free
(To\'{t}h 2000), and Noble's 2D primitive variable inversion method
(Noble et al. 2006).

We consider the Riemann problem consisting of two uniform initial
states (a left- and a right state) with different and discontinuous
hydrodynamic properties specified by the rest-mass density $\rho$, the
gas pressure $p$, the specific internal energy $u$, the specific
enthalpy $h \equiv 1+ u/c^{2} + p/\rho c^{2}$, and with velocity
component $v^{t}=v^{z}$ (the jet-direction) tangent to the initial
discontinuity. We consider the right state (the medium external to the
jet) to be a ``colder" fluid with a large rest-mass density and
essentially at rest. Specifically, we select the following initial
conditions: $\rho_{R}=10^{-2} \rho_{0}$, $p_{R}=1.0 \rho_{0} c^{2}$,
$v^{n}_{R}=v^{x}_{R}=0.0$, and $v^{t}_{R} =v^{z}_{R}=0.0$, where
$\rho_{0}$ is an arbitrary normalization constant (our simulations are
scale-fee) and $c$ is the speed of light in vacuum, $c=1$. The left
state (jet region) is assumed to have lower density, higher
temperature and higher pressure than the colder, denser right state,
and to have a relativistic velocity tangent to the discontinuity
surface. Specifically, $\rho_{L}=10^{-4} \rho_{0}$, $p_{L}=10.0
\rho_{0} c^{2}$, $v^{n}_{L}=v^{n}_{L}=0.0$, and
$v^{t}_{L}=v^{z}_{L}=0.99c$ ($\gamma_{L} \simeq 7$) (in Table 1 these
conditions are collectively labeled as case HDA). 
The fluid satisfies an adiabatic $\Gamma-law$ equation of
state with $\Gamma=4/3$.  
The relevant sound speeds are $a_{j}=0.57735c$ in the jet flow, and 
$a_{e} = 0.57663c$ in the external medium, where the sound speed is 
given by $a=\sqrt{ \Gamma p/\rho h }$.  
We note that if the adiabatic index were $\Gamma=5/3$, the sound
speeds would be $a_{j,e} \sim 0.82c$. These velocities exceed the
maximum physically allowed sound speed $a = c/\sqrt{3}$. 
Therefore we choose the
adiabatic index to be $\Gamma=4/3$ in our simulations.
Figure 1 shows a schematic depiction of the geometry of our simulations.

To investigate the effect of magnetic fields, we consider the
following left state field geometries: ``poloidal'', $B^{z}=6.0
\sqrt{\rho_{0} c^{2}}$ ($B^{'}_{z} =6.0 \sqrt{\rho_{0} c^{2}}$), in
the MHDA case, and ``toroidal'' (not truly toroidal but we use this
designation for simplicity), $B^{y}=42.0 \sqrt{\rho_{0} c^{2}}$
($B^{'}_{y}=6.0 \sqrt{\rho_{0} c^{2}}$), in the MHDB case (see
Table1), where $B^{'}_{i}$ is the magnetic field measured in the jet
fluid frame ($B^{'}_{y}=B^{y}/\gamma$, $B^{'}_{z}=B^{z}$). Although
the strength of the magnetic field measured in the laboratory frame
($B^{i}$) in the left state is larger in the MHDB case than the MHDA
case, the magnetic pressure ($p_{mag}$) is the same as that of the
MHDA case ($p_{mag}=(B^{'})^{2}/2$). 
The relevant Alfv\'{e}n speed in the left state is 
$v_{Aj}=0.68825c$, whereas the Alfv\'{e}n speed $v_{A}$ is given 
by $v_{A}=\sqrt{[(B^{'})^{2}/c^{2}]/[ \rho h + (B^{'})^{2}/c^{2}]}$.

For comparison, the HDB case listed in Table 1 is a high gas pressure,
pure-hydrodynamic case ($p_{L}=28.0 \rho_{0}c^{2}$). In this case the
gas pressure $p_{L}$ in left state is equal to the total pressure
($p_{tot}$) in the MHD cases ($p_{tot}=p_{gas}+p_{mag}$) in the left
state.   

We employ free boundary conditions in
all-directions. The simulations are performed in the region $-0.2 \le
x \le 0.2$ with 6400 computational zones ($\Delta x = 6.25
\times 10^{-5}$) until simulation time $t=0.1$.  We emphasize that
our simulations are scale-fee. If we specify a system of size
$L=10^{7} $ cm ($\Delta L \simeq 6.25 \times 10^{2}$ cm), a simulation
time of $t=0.2$ corresponds to about $0.06$ msec. The units of
magnetic field strength and pressure depend on the normalization of
the density. If we take, for example, the density unit to be
$\rho_{0}=10^{-20} \, \rm{g\, cm}^{-3}$, the magnetic field strength
unit is about $3$ G and the pressure unit is $P \simeq 10\, \rm{dyn \,
cm}^{-2}$.

\section{Results}
\subsection{Effects of the magnetic field in 1-D simulations}

Figure 2 shows the radial profiles of density, gas pressure, velocity
normal to the interface ($v^{x}$) - hereafter normal velocity - and
velocity tangent to the interface ($v^{z}$)- hereafter tangential
velocity - for case HDA. The solution displays a right-moving shock, a
right-moving contact discontinuity and a left-moving rarefaction wave
($_{\leftarrow}RCS_{\rightarrow}$). This hydrodynamical profile is
similar to that found by Rezzolla et al. (2003) and Aloy \& Rezzolla
(2006). The simulation results (dashed lines) are in good agreement
with the exact solution (solid lines, calculated with the code of
Giacomazzo \& Rezzolla 2006) except for the spike in the normal
velocity $v^{x}$. Otherwise the normal velocity and propagation of the
shock propagating to the right (the forward shock) is $v^{x}
\sim 0.082c$ where this value is
determined from the exact solution.  The small spike evident in
Fig. 2 is a numerical artifact and is seen in all simulation results
(e.g., in the middle panel of Fig.\ 3) at the right moving shock
($S_{\rightarrow}$). This numerical spike is reduced by higher
resolution calculations (see Appendix A). In the left-moving
rarefaction ($_{\leftarrow}R$) region, the tangential velocity
increases as a result of the hydrodynamical boosting mechanism
described by Aloy \& Rezzolla (2006). In the case shown in Figure 2
the jet is accelerated to $\gamma \sim 12$ from an initial Lorentz
factor of $\gamma_{L} \simeq 7$.

Figure 3 displays the resulting profiles of gas pressure, normal
velocity ($v^{x}$) and tangential velocity ($v^{z}$) of
the magnetohydrodynamic cases MHDA (blue), MHDB (red), and the
high pressure, pure-hydrodynamic case HDB (green).
In the magnetohydrodynamic cases, the magnetization parameter
$\sigma \equiv (B^{'})^{2}/\rho h$ and the plasma beta parameter $\beta
\equiv p_{gas}/p_{mag}$ (on the left side) are 0.556 and 0.45,
respectively. The resulting structure consists of a
right-propagating fast shock, a right-propagating contact discontinuity,
and a left-propagating fast rarefaction wave
($_{\leftarrow}R_{F}CS_{F \rightarrow}$).

In the MHDA case ($B^{z}=6.0$ ($B^{'}_{z}=6.0$)) shown as blue curves, the
right-moving fast shock ($S_{F \rightarrow}$) and the left-moving fast
rarefaction wave ($_{\leftarrow}R_{F}$) are
stronger than the related structures in the HDA case. Consequently,
the normal velocity ($v^{x} \sim 0.172c$) is larger than that
for the HDA case ($v^{x} \sim 0.082c$). The tangential velocity
($v^{z} \sim 0.9915c$) is lower than that of the HDA case ($v^{z}
\sim 0.9933c$). These velocity values are determined from the exact
solution. Although the acceleration in the z-direction is weaker,
the jet experiences a larger total acceleration than in the HDA case due
to the larger normal velocity, and the jet Lorentz factor reaches
$\gamma \sim 15$. Thus the ``poloidal'' magnetic field in the jet
region strongly affects sideways expansion, shock profile and total
acceleration.

In the MHDB case ($B^{y}=42.0$ ($B^{'}_{y}=6.0$)) shown as red curves,
the right-moving fast shock ($S_{F \rightarrow}$) is slightly weaker
than in the HDA case, and the resulting normal velocity ($v^{x} \sim
0.080c$) is slightly less than in the HDA case ($v^{x} \sim
0.082c$). The left-propagating fast rarefaction wave
($_{\leftarrow}R_{F}$) is stronger than what we found for the
HDA case. Therefore the tangential velocity ($v^{z} \sim 0.9958c$) is
higher than in the HDA case ($v^{z} \sim 0.9933c$). These velocity values 
are determined from the exact solution. Altough 
the ``toroidal'' magnetic field in the jet region does not greatly
affect the sideways expansion and shock profile, the resulting
total acceleration to $\gamma \sim 19$ is larger than in
the HDA case.

To investigate the effect of the total pressure, we performed a
pure-hydrodynamic simulation with high gas pressure (case HDB),
shown as green curves, equal to the total (gas plus magnetic)
pressure ($p_{tot} =p_{gas}+p_{mag}$) in the MHD cases. The
resulting structure for this case is the same as that of HDA case
($_{\leftarrow}RCS_{\rightarrow}$). The right-moving shock
($S_{\rightarrow}$) and the left-moving rarefaction wave
($_{\leftarrow}R$) are slightly stronger than those in the HDA case
because of the initial high gas pressure in left state.
Consequently, the normal velocity $v^{x}$ in the HDB case is larger
($v^{x} \sim 0.108c$) than in the HDA case ($v^{x} \sim 0.0822c$).
In the region of the left-propagating rarefaction wave
($_{\leftarrow}R$), the tangential velocity is the same as that 
in the HDA case, the jet accelerates only with a marginally greater
efficiency than in the HDA case, and the resulting Lorentz factor
thus reaches only $\gamma \sim 15$.

Although the total pressure is the same in the hydro HDB case and MHD
cases, the existence and direction of the magnetic field changes the
shock profiles and acceleration. We summarize the acceleration
properties of the different cases in Table 2 where velocity
values are determined from the exact solutions. When the gas
pressure becomes large in the left state, the normal velocity increases and 
the jet is more efficiently accelerated.
 This is because the larger discontinuity in the gas pressure produces a
stronger forward shock as well as stronger rarefaction. In MHD, the
magnetic pressure is measured in the jet fluid frame and depends on
the angle between the flow and magnetic field.  The magnetosonic
speeds also depend on the angle between the flow and the magnetic
field, even for the same magnetic pressure.  The direction of the
magnetic field is thus a very important geometric parameter for
relativistic magnetohydrodynamics.  When a ``poloidal'' magnetic field
($B^{z}$) is present in the jet region, larger sideways expansion is
produced, and the jet can achieve higher speed due to the contribution
from the normal velocity.
By contrast, when a ``toroidal'' magnetic
field ($B^{y}$) is present in the jet region, although the shock profile is
only changed slightly, the jet is more accelerated in
the tangential direction due to the additional contribution of the
tangential component of the Lorentz force
($\mathbf{F}_{EM,z}=(\mathbf{J} \times \mathbf{B})_{z}$) shown in
Figure 4 (in the MHDA case there is no additional force).
It should be
noted that the region with high Lorentz force is approximately
coincident with the acceleration region $-0.025$ to $0.0$ and 
the force still exists at time $t=0.2$. The region with the 
highest Lorentz force is at the inner edge of the acceleration 
region.
From an efficiency point of view, a ``toroidal'' magnetic field
with the same strength in the jet fluid frame and the same
magnetic pressure as those of a ``poloidal'' field provides the 
most efficient acceleration.  A ``poloidal'' magnetic
field  provides
acceleration comparable to that resulting from high gas pressure,
e.g., the HDB case.

\subsection{Dependence of the MHD boost mechanism on magnetic field strength}

To investigate the acceleration efficiency of the magnetic field, we
compare jet speeds for the MHDA and MHDB cases and the results are
shown in Figure 5.  The left panels in Figure 5 show the dependence of
the maximum tangential and normal velocities and resulting Lorentz
factors on the strength of the poloidal ($B^{'}_{z}$) component of the
magnetic field in the fluid frame. The solid line indicates values
obtained using the code of Giacomazzo \& Rezzolla (2006) and the
crosses indicate values obtained from our simulations at time
$t=0.2$. For numerical reasons our code does not yield a solution for
$B^{'}_{z} > 10$ ($B^{z} > 10$) (the simulation results are indicated by
the crosses). When the poloidal magnetic field increases, the code of 
Giacomazzo \& Rezzolla (2006) indicates that the maximum normal velocity 
increases and the maximum tangential velocity deceases. The break near
$B^{'}_{z} \simeq 4$ occurs near the transition (in the left state)
from gas pressure dominated to magnetic pressure dominated.  The
Lorentz factor results shown in Figure 5 (left bottom panel) indicate
that a sufficiently strong poloidal magnetic field in the jet region
will allow a jet to achieve $\gamma \sim 22$, even if the jet is only
``mildly'' relativistic initially, i.e., $\gamma_{L} \sim 7$. We
note that in the hydrodynamic cases investigated by Aloy \& Rezzolla
(2006) the Lorentz factor decreases as the normal velocity increases
(see their Fig.\ 4) where we find that the Lorentz factor increases as
the normal velocity increases. Our result is different from that of
Aloy \& Rezzolla (2006) because their initial conditions were
different.  In particular, they varied the initial normal velocity in
the jet region while holding the initial Lorentz factor constant.

The right panels in Figure 5 show the dependencies of the maximum
tangential and normal velocities and the Lorentz factor on the
strength of the toroidal ($B^{'}_{y}$) component of the magnetic field
as measured in the jet fluid frame. Again, the solid line indicates
values obtained with the code of Giacomazzo \& Rezzolla (2006) and the
crosses indicate values obtained from our simulations at time
$t=0.2$. 
When the toroidal magnetic field becomes large in the jet
region, the maximum normal velocity increases initially, then decreases when $B^{'}_{y} > 4$, and the maximum
tangential velocity increases. This dependence is opposite to that of
the poloidal magnetic field. The acceleration in the tangential
direction occurs due to the additional contribution of the Lorentz
force shown in Figure 4. When the toroidal magnetic field becomes
large in the jet region, the Lorentz force in the tangential direction
increases and contributes to a large acceleration of the jet in the
tangential direction. The transition from gas pressure dominated to
magnetic pressure dominated left states occurs near $B^{'}_{y} \simeq
4$. 
This change from gas to magnetic pressure dominated is reflected
 in the normal velocity profile. The acceleration is much larger than that
found in the comparable poloidal magnetic field case. While at
$B^{'}_{y} \simeq 20$ the maximum Lorentz factor reaches $\gamma
\sim 72$, at $B^{'}_{z} \simeq 20$ the maximum Lorentz factor is
only $\gamma \sim 22$.

\subsection{Multidimensional simulations}

To investigate the effects induced by more than one degree of freedom,
we perform two dimensional RMHD simulations of the MHDA case
($B^{'}_{z}=6.0$). The computational domain corresponds to a local
part of the jet flow. In the simulations, a ``pre-existing'' jet flow
is established across the computational domain. The initial condition
is the same as that of the 1D MHDA case (e.g., $v^{z}=0.99c$ and
$B^{'}_{z}=6.0$). In order to investigate a possible influence of the
chosen coordinate system, we perform the calculations in Cartesian and
cylindrical coordinates. The discontinuities between the jet and the
external medium are set at $x$ or $r=1.0$ in the initial state (see
Fig. 1). The computational domain is $0.5 \le x, r \le 1.5$ and $0 \le
z \le 5.0$ with $(N_{x, r} \times N_{z}) = (2000 \times 250)$, where
$N_{x, r}$ and $N_{z}$ are the number of computational zones in the
$x$ or $r$ direction and in the $z$ direction. We use a large number
of computational zones in the $x$ or $r$ direction in order to
satisfactorily resolve the shock profile. We impose periodic boundary
conditions in the z-direction and free boundary conditions in the
$x$ or $r$ direction.
The computational domain is far from the jet center in order to
obtain high resolution near the jet surface.  In this case it is
necessary to use free boundary conditions at the inner boundary in $x$
or $r$.  Here we are far from the jet axis and waves and fluid must be
free to move towards the axis through this boundary and not experience a
reflection.

The initial condition for these 2D simulations is a simple
extension of the 1D MHDA case into the z-direction and represents the
temporal development of a planar (Cartesian coordinates) or
cylindrical interface that is infinite in extent in the z-direction.
Effectively we consider a local part of a jet flow.  Here we consider
only the poloidal magnetic field case as a uniformly overpressured
cylindrical jet containing a uniform poloidal magnetic field as valid
physically.  The ``toroidal'' magnetic field case in 1D cannot be
compared to a proper cylindrical toroidal magnetic field in which hoop
stresses and radial gradients will play a role.

Figure 6 shows 2D images of the Lorentz factor for the 2D MHDA
simulation in Cartesian and in cylindrical coordinates at time
$t=0.6$. The left-moving rarefaction waves do not reach the inner
boundary ($x$ or $r$ direction) so the choice of inner outflow
boundary condition does not influence the results. In both cases, a
thin surface is accelerated by the MHD boost mechanism to reach a
maximum Lorentz factor $\gamma \simeq 15$ from an initial Lorentz
factor $\gamma_{L}
\simeq 7$. The jet in cylindrical coordinates is slightly more
accelerated than the jet in Cartesian coordinates.  The presence of
velocity shear between the jet and external medium can excite
Kelvin-Helmholtz (KH) instabilities (e.g., Ferrari et al. 1978; Hardee
1979, 1987, 2000; Birkinshaw 1984), and such instability might affect
the relativistic boost mechanism. However, we do not see any growth
of the KH instability during the simulation. This is because the
simulation duration is too short for KH instabilities to grow. However,
in longer duration RMHD cylindrical jet simulations KH instabilities
can grow (Hardee 2007; Mizuno et al.\ 2007) and this might effect 
significantly the later stages of jet evolution.

In order to investigate simulation results quantitatively, we have
taken one-dimensional cuts through the computational box perpendicular
to the z-axis.  Figure 7 shows the resulting profiles of gas pressure,
Lorentz factor ($\gamma$), normal velocity ($v^{x}$ or $v^{r}$) and
tangential velocity ($v^{z}$) of the 2D MHDA cases in Cartesian
(dotted lines) and in cylindrical coordinates (dashed lines). The
exact solution of the 1D MHDA case is shown as solid lines.  The
result consists of a right-moving fast shock, right-moving contact
discontinuity, and a left-moving fast rarefaction wave
($_{\leftarrow}R_{F}CS_{F \rightarrow}$). The profiles from the 2D
MHDA simulation in Cartesian coordinates match well those of the 1D
MHDA case.  In the 2D MHDA simulation with cylindrical coordinates,
the right-moving fast shock ($S_{F \rightarrow}$) is weaker and the
left-moving fast rarefaction wave ($_{\leftarrow}R_{F}$) is slightly
stronger than those of the 2D MHDA simulation with Cartesian
coordinates.  Selecting cylindrical coordinates, causes the normal
velocity to decrease gradually in the expansion. The tangential
velocity in cylindrical coordinates ($v^{z} \sim 0.991530c$) is
slightly faster than in Cartesian coordinates ($v^{z} \sim 0.991500c$).
Thus the jet Lorentz factor reaches $\gamma \sim 16$ in cylindrical
coordinates and $\gamma \sim 15$ in Cartesian coordinates.  This
result suggests that different coordinate systems affect sideways
expansion, shock profile, and acceleration slightly.

\section{Summary and Discussion}

We performed relativistic magnetohydrodynamic simulations of an
acceleration boosting mechanism for fast astrophysical jet flows that
results from highly overpressured, tenuous flows with an initially
modest relativistic speed relative to a colder, denser external medium
at rest.  We employed the RAISHIN code (Mizuno et
al. 2006a), to study the relativistic boost mechanism proposed by Aloy
\& Rezzolla (2006), who showed that hydrodynamic accelerations to
$\gamma > 1000$ are possible in the situation described above.  For
numerical reasons, we reduced the pressure discontinuity between the
hotter higher pressure jet and colder lower pressure external medium
and also reduced the initial jet velocity (see Apendix A3).  
Our results still show the
same behavior ($_{\leftarrow}RCS_{\rightarrow}$) found in Rezzolla et
al.\ (2003) and Aloy \& Rezzolla (2006). The same hydrodynamical
structures emerge in our simulation, confirming the basic properties
of the boost mechanism proposed in their work. Subsequently we
extended their investigation to study the effects of magnetic fields
that are parallel (poloidal) and perpendicular (toroidal) to the flow
direction but parallel to the interface.

Our simulations show that the presence of a magnetic field in the jet
can significantly change the properties of the outward moving shock
and inward moving rarefaction wave, and can in fact result in even
more efficient acceleration of the jet than in a pure-hydrodynamic
case. In particular, the presence of a toroidal magnetic field
perpendicular to jet flow produces a stronger inward moving rarefaction 
wave. This leads to acceleration from
$\gamma \sim 7$ to $\gamma \gtrsim 15$ when the magnetic pressure is
comparable to the gas pressure. A comparable pure-hydrodynamic case
yields acceleration to $\gamma \lesssim 12$. Our results would
indicate acceleration to $\gamma \sim 100$ for a case with magnetic 
pressure 40
times the gas pressure. Thus, the magnetic field can in principle play
an important role in this relativistic boost mechanism.

We found that a jet with a flow aligned poloidal field was slightly
more accelerated in cylindrical coordinates than one in Cartesian
coordinates but in general our 1D and 2D results for the poloidal
field appear comparable.  The current simple 2D MHD simulation in
cylindrical coordinates is directly applicable to a 3D cylindrical
geometry where the magentic field and jet flow are aligned and tangent
to the jet-external medium interface. However, recent GRMHD
simulations of jet formation predict that the jet has a rotational
velocity and considerable radial structure (e.g., Nishikawa et
al. 2005; Mizuno et al. 2006b; De Villiers et al. 2005; Hawley \&
Krolik 2006; McKinney \& Gammie 2004; McKinney 2006).  The effect of
such radial structure on this boost mechanism is yet to be determined.

Our present results for the 1D ``toroidal'' field are not likely to
apply in a 3D cylindrical geometry where a toroidal field exerts a
hoop stress that does not exist in the 1D configuration. It seems
likely that this hoop stress would so modify the sideways expansion of
an overpressured cylindrical jet as to render our present toroidal
field results not applicable unless the magnetic field in 3D is
tangled or the thin boost region is relatively insensitive to radial
gradients.  In order to properly investigate full 3D effects, it will
be necessary to perform full 3D RMHD simulations including toroidal
and helical magnetic fields.

The initial conditions in our present 2D simulations are a simple
extension of the 1D poloidal MHD case, models a local part of an
overpressured jet flow in a colder denser ambient and provides a
first step toward multi dimensional simulations. To address the
question whether or not such strong, magnetically enhanced boosts
really do take place in astrophysical sources (AGNs, quasars,
microquasars, gamma-ray bursts) will require additional numerical
simulations to show that this process can work for jets injected into
a reasonable astrophysical environment.
The operation of the MHD boost is likely to be strongly affected by
the properties of the external medium, expected jet overpressures, and
spatial development of the jet flow and external medium downstream
from the jet source. 
For example, it is conceivable that magnetic
pressure effects are more dominant relative to thermal pressure
effects in AGN jets where a magnetically dominated ``Poynting''
flux jet is confined by a colder, denser external medium.
 A hot GRB fireball can expand
and accelerate under its thermal pressure to reach large Lorentz
factors as long as baryon-loading is small (M\'esz\'aros et al. 1993;
Piran et al. 1993). Although this simple model can account for the
large ($>100$) Lorentz factors inferred for GRBs, it does not reflect
more realistic settings of complex GRB progenitor/central engine
models. In the collapsar model for long-duration GRBs (Woosley 1993),
the tenuous jet is believed to propagate in a surrounding dense
stellar envelope (Zhang et al. 2003), so that the hydrodynamic
configuration considered by Aloy \& Rezzolla (2006) and in this paper
is naturally satisfied. A strong poloidal magnetic field is likely
present at the central engine. 
In some GRB models, the flow is even dominated by a Poynting flux 
(Lyutikov 2006).
In this case the magnetohydrodynamic
boost mechanism discussed here would then play an important role in
jet acceleration. The final Lorentz factor should depend on the
detailed parameters invoked in this mechanism as well as the unknown
baryon loading process during the propagation of the jet in the
envelope. In the case of short GRBs that may be of compact star merger
origin (e.g., Pacz\'ynski 1986; Nakar 2007), there is no dense stellar 
envelope
surrounding the jet. The jet region is nonetheless more tenuous than
the surrounding medium due to the centrifugal barrier in the jet, so
that the acceleration mechanism discussed here still applies (e.g.,
see Aloy et al. 2005 for the pure hydrodynamic case). Due to a likely
smaller baryon loading in the merger environment, the jet may achieve
an even higher Lorentz factor than for the case of long GRBs, as
suggested by some observations (e.g. their harder spectrum and shorter
spectral lags). The magnetohydrodynamic acceleration mechanism
discussed here also naturally yields a GRB jet with substantial
angular structure. In particular, since acceleration is favored in the
rarefaction region near the contact discontinuity, this mechanism
naturally gives rise to the kind of ring-shaped jet that has been
discussed in some empirical GRB models (e.g. Eichler \& Levinson
2006).

\appendix
\section{Resolution Tests}
\subsection{One-dimensional shock tube tests with transverse velocity}

Recently it has been reported that it is numerically challenging to resolve
1D shock tube test problems with transverse velocities using the same
number of computational zones used in the absence of transverse
velocities (Mignone \& Bodo 2005; Zhang \& MacFadyen 2006; Mizuta et
al.\ 2006). We have performed a resolution study using from 800 to
12,800 uniform zones spanning $L_{x} = 1.0$, where $L_{x}$ is the
simulation size in the x direction, and we have used the initial
conditions from Mignone \& Bodo (2005) and Mizuta et al.\ (2006) with
adiabatic index $\Gamma = 5/3$ as follows:

\noindent
Left state $(0 < x < 0.5)$: $\rho_{L}=1.0$, $p_{L}=1.0 \times 10^3$,
$v^{x}_{L}=0$, $v^{z}_{L} = 0.9~c$. 

\noindent
Right state $(0.5 < x < 1.0)$:
$\rho_{R}=1.0$, $p_{R}=1.0 \times 10^{-2}$, $v^{x}_{R}=0$,
$v^{z}_{R}=0$.

The results of numerical and analytical solutions to this test problem
are shown in Figure 8. The left-going rarefaction ($_{\leftarrow}R$)
is resolved with good accuracy even at lower resolution. On the
contrary, both right-going shock and contact discontinuities
($CS_{\rightarrow}$) are not resolved in both position and value in
lower resolution calculations. In calculations with higher resolution
(typically $N_{x} \ge 6400$ over $L_{x}=1.0$) the shock front position
is calculated correctly. However, there still remains some undershoot
in $v^{z}$ at the contact discontinuity.

\subsection{One dimensional hydrodynamic relativistic boost model}

We also performed a resolution study of the HDA case. The initial 
condition is described in Table 1. The simulations have been performed 
using from 800 to 12,800 uniform zones spanning $L_{x} = 0.4$ equivalent 
to using 2000 to 32,000 uniform zones spanning $L_{x} = 1.0$.

The results of numerical and analytical solutions are shown in Figure
9. The left-going rarefaction ($_{\leftarrow}R$) is resolved to good
accuracy at even the lowest resolution. On the other hand, the
right-going shock ($S_{\rightarrow}$) is not sufficiently resolved in
both position and value in lower resolution calculations. Higher
resolution calculations ($N_{x} \ge 6400$) determine the shock front
position with good accuracy. However, there still remains some
overshoot in $v^{x}$ behind the right-going shock.

We conclude from this study that the use of 6400 computational zones
spanning $L_{x} = 0.4$ used for our 1D simulations provides excellent
quantitative accuracy.  The use of 2000 computational
zones spanning $L_{x} = 1.0$, equivalent to 800 zones spanning $L_{x}
= 0.4$, provides sufficient quantitative accuracy for the comparison
between 1D and 2D results.

\subsection{One dimensional hydrodynamic 
relativistic boost model (Aloy \& Rezzolla model)}

We have performed a resolution study of the one dimensional hydrodynamic relativistic boost model proposed by Aloy \& Rezzolla (2006). The simulations have been performed using from 800 to 12,800 uniform zones spanning $L_{x}=0.4$. 
Here we have used the initial density and pressure conditions in the left and right states with adiabatic index $\Gamma = 5/3$ from Aloy \& Rezzolla (2006) but with a reduced tangential velocity as follows:

\noindent
Left state $(-0.2 < x < 0.0)$: $\rho_{L}=1.0 \times 10^{-4}$, $p_{L}=1.0 \times 10^{-3}$,
$v^{x}_{L}=0$, $v^{z}_{L} = 0.99~c$. 

\noindent
Right state $(0.0 < x < 0.2)$:
$\rho_{R}=1.0 \times 10^{-2}$, $p_{R}=1.0 \times 10^{-6}$, $v^{x}_{R}=0$,
$v^{z}_{R}=0$.

The results of numerical and analytical solutions are shown in Figure 10. 
The left-going rarefaction ($_{\leftarrow}R$), right-going contact discontinuity and shock ($CS_{\rightarrow}$) are not sufficiently resolved in both position and value in lower resolution calculations. Higher resolution calculations ($N_{x} > 6400$) determine the position and value of the left-going rarefaction wave with good accuracy. However, even in the highest resolution calculation ($N_{x}=12,800$) still the right-going contact discontinuity and shock are not 
resolved. 
We conclude from this study that in order to resolve the Aloy \& Rezzolla model with good accuracy we would have to perform simulations with much higher resolution and employ the Adaptive Mesh Refinement method like that used by Zhang \& MacFadyen (2006) to obtain locally higher resolution at the area where shocks and contact discontinuities exist to save CPU time and memory.

\acknowledgments

Y. M. is supported by an appointment of the NASA Postdoctoral
Program at NASA Marshall Space Flight Center, administered by Oak
Ridge Associated Universities through a contract with NASA. P. H.
acknowledges partial support by National Space Science and
Technology Center (NSSTC/NASA) cooperative agreement NCC8-256 and
NSF award AST-0506666 to the University of Alabama. K. N.
acknowledges partial support by NSF awards ATM-0100997, INT-9981508,
and AST-0506719, and the NASA award NNG05GK73G, HST-AR-10966.01-A,
and NASA-06-SWIFT306-0027 to the University of Alabama in
Huntsville. B. Z. acknowledges partial support by NASA award
NNG05GB67G and NNG06GH62G. The simulations have been performed on
the IBM p690 (copper) at the National Center for Supercomputing Applications
(NCSA) which is supported by the NSF and Altix3700 BX2 at YITP in
Kyoto University.

\begin{deluxetable}{llcccccccc}
%\tabletypesize{\small}
\tablecolumns{10}
\tablewidth{0pc}
\tablecaption{Model and Parameters \label{table1}}
\tablehead{
\colhead{Case}&  & \colhead{$\rho$} & \colhead{$p$} & \colhead{$v^{x}$} &
\colhead{$v^{y}$} & \colhead{$v^{z}$} & \colhead{$B^{x} (B^{'}_{x})$} & \colhead{$B^{y} (B^{'}_{y})$} & \colhead{$B^{z} (B^{'}_{z})$} }
\startdata
\textbf{HDA} & {\itshape left state\/} & $10^{-4}$ & 10.0
& 0.0 & 0.0 & 0.99 & 0.0(0.0) & 0.0(0.0) & 0.0(0.0) \\
             & {\itshape right state\/} & $10^{-2}$ & 1.0
& 0.0 & 0.0 & 0.0 & 0.0(0.0) & 0.0(0.0) & 0.0(0.0) \\
 \hline
\textbf{HDB} & {\itshape left state\/} & $10^{-4}$ & 28.0
& 0.0 & 0.0 & 0.99 & 0.0(0.0) & 0.0(0.0) & 0.0(0.0) \\
             & {\itshape right state\/} & $10^{-2}$ & 1.0
& 0.0 & 0.0 & 0.0 & 0.0(0.0) & 0.0(0.0) & 0.0(0.0) \\
 \hline
 \textbf{MHDA} & {\itshape left state\/} & $10^{-4}$ & 10.0
& 0.0 & 0.0 & 0.99 & 0.0(0.0) & 0.0(0.0) & 6.0(6.0) \\
             & {\itshape right state\/} & $10^{-2}$ & 1.0
& 0.0 & 0.0 & 0.0 & 0.0(0.0) & 0.0(0.0) & 0.0(0.0) \\
 \hline
 \textbf{MHDB} & {\itshape left state\/} & $10^{-4}$ & 10.0
& 0.0 & 0.0 & 0.99 & 0.0(0.0) & 42.0(6.0) & 0.0(0.0) \\
             & {\itshape right state\/} & $10^{-2}$ & 1.0
& 0.0 & 0.0 & 0.0 & 0.0(0.0) & 0.0(0.0) & 0.0(0.0) \\

\enddata
\tablecomments{HDA is a hydrodynamic case. MHDA and MHDB are
magnetohydrodynamic cases with $B^{z}_{L}=6.0$ ($B^{'}_{z, L}=6.0$)
and $B^{y}_{L}=42.0$ ($B^{'}_{y, L}=6.0$) respectively. HDB is a
hydrodynamic case with gas pressure $p_{gas, L}$ equal to $p_{tot,
L}=p_{gas, L}+p_{mag, L}$ in the MHD cases.}
\end{deluxetable}

\newpage

\begin{figure}
\epsscale{0.5}
\plotone{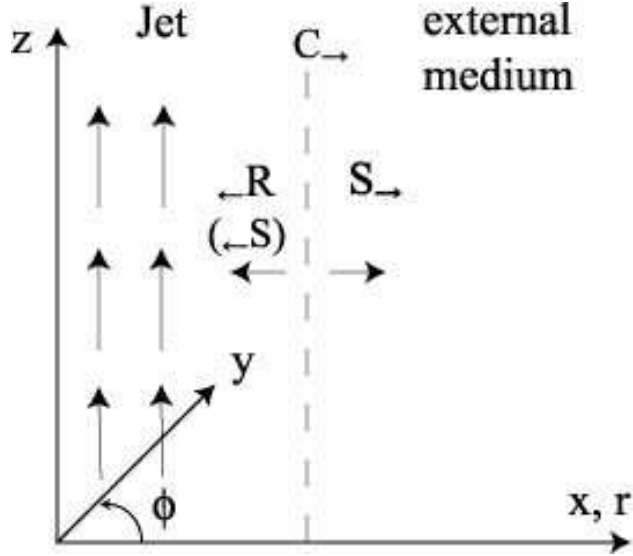}
\caption{Schematic picture of our simulations. \label{f1}}
\end{figure}

\begin{figure}
\epsscale{0.7}
\plotone{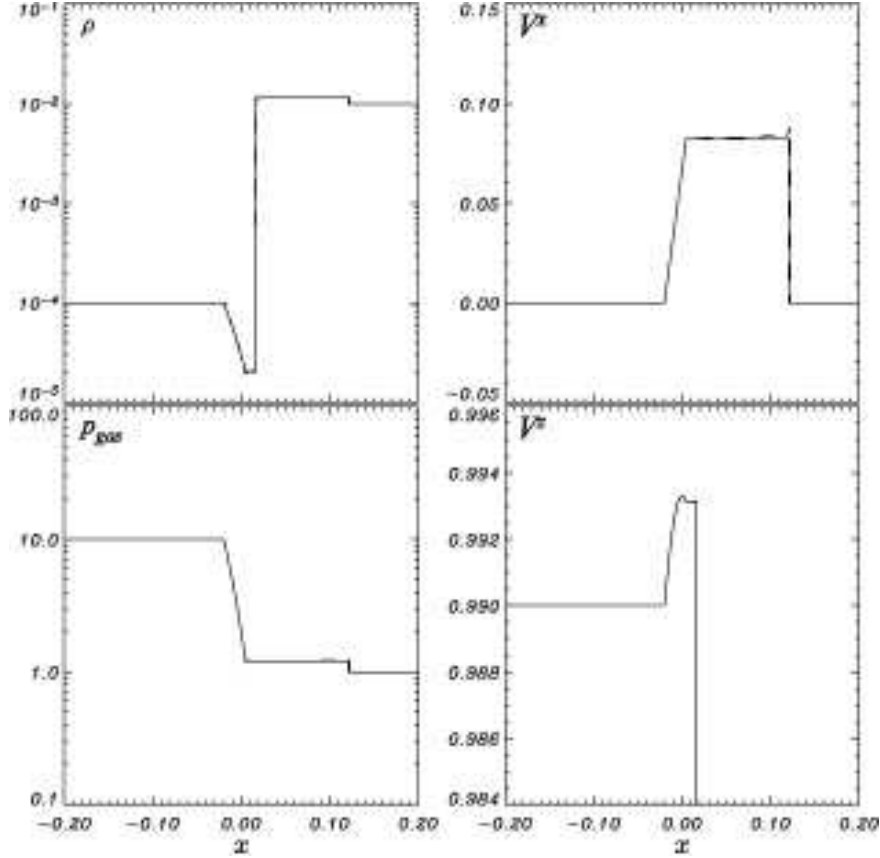}
\caption{Profiles of ({\it left-upper panel}) density, ({\it
left-lower panel}) gas pressure, ({\it right-upper panel}) normal
velocity ($v^{x}$), and ({\it right-lower panel}) tangential velocity
($v^{z}$) in the HDA case at time $t=0.2$. The solid lines are the exact
solution and the dashed lines are the simulation results. 
\label{f2}}
\end{figure}

\begin{figure}
\epsscale{0.3}
\plotone{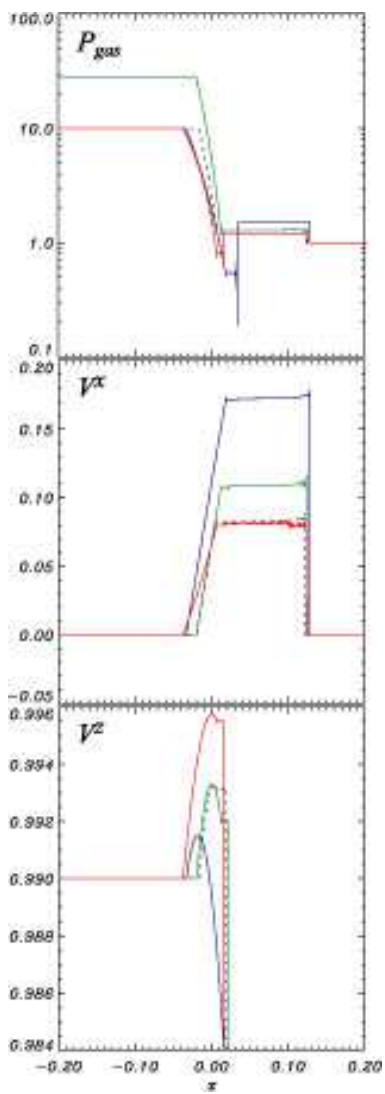}
\caption{Profiles of gas pressure ({\it upper panel}), normal velocity
($v^{x}$) ({\it middle panel}), and tangential velocity ($v^{z}$)
({\it lower panel}) in the HDB (green), MHDA (blue), MHDB (red), and HDA
(dotted-line) cases at time $t=0.2$.
\label{f3}}
\end{figure}

\newpage

\begin{deluxetable}{lcccccccc}
%\tabletypesize{\small}
\tablecolumns{4} \tablewidth{0pc}
\tablecaption{Maximum velocities and Lorentz factor \label{table2}}
\tablehead{ \colhead{Case} & \colhead{$v^{x}$} & \colhead{$v^{z}$} &
\colhead{$\gamma$} }
\startdata
\textbf{HDA} & 0.082211 & 0.993292 & 11.9820 \\
\textbf{HDB} (high gas pressure)& 0.107550 & 0.993293 & 15.2890 \\
\textbf{MHDA} (poloidal field) & 0.171533 & 0.991502 & 14.8255 \\
\textbf{MHDB} (toroidal field) & 0.080350 & 0.995750 & 19.9020 \\
\enddata
\tablecomments{
Velocity values determied from exact solutions.
}
\end{deluxetable}

\begin{figure}
\epsscale{0.4} 
\plotone{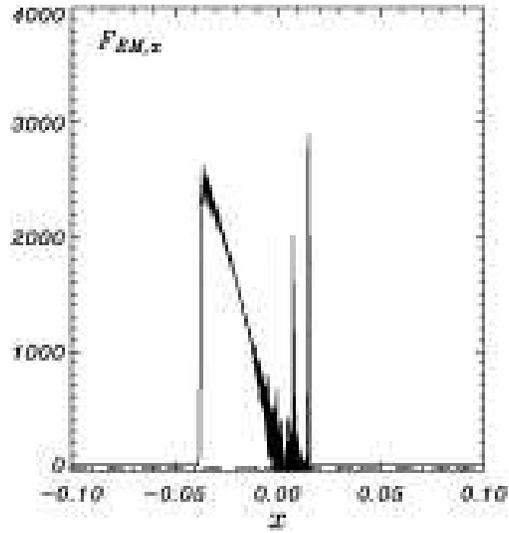}
\caption{Profile of tangential component of Lorentz force ($F_{EM,z}=(\mathbf{J}
\times \mathbf{B})_{z}$) of the MHDA ($B^{'}_{z}=6.0$)({\it dashed-line}) and the MHDB ($B^{'}_{y}=6.0$)({\it solid-line}) cases at time $t=0.2$.
\label{f3-1}}
\end{figure}

\begin{figure}
\epsscale{0.8}
\plotone{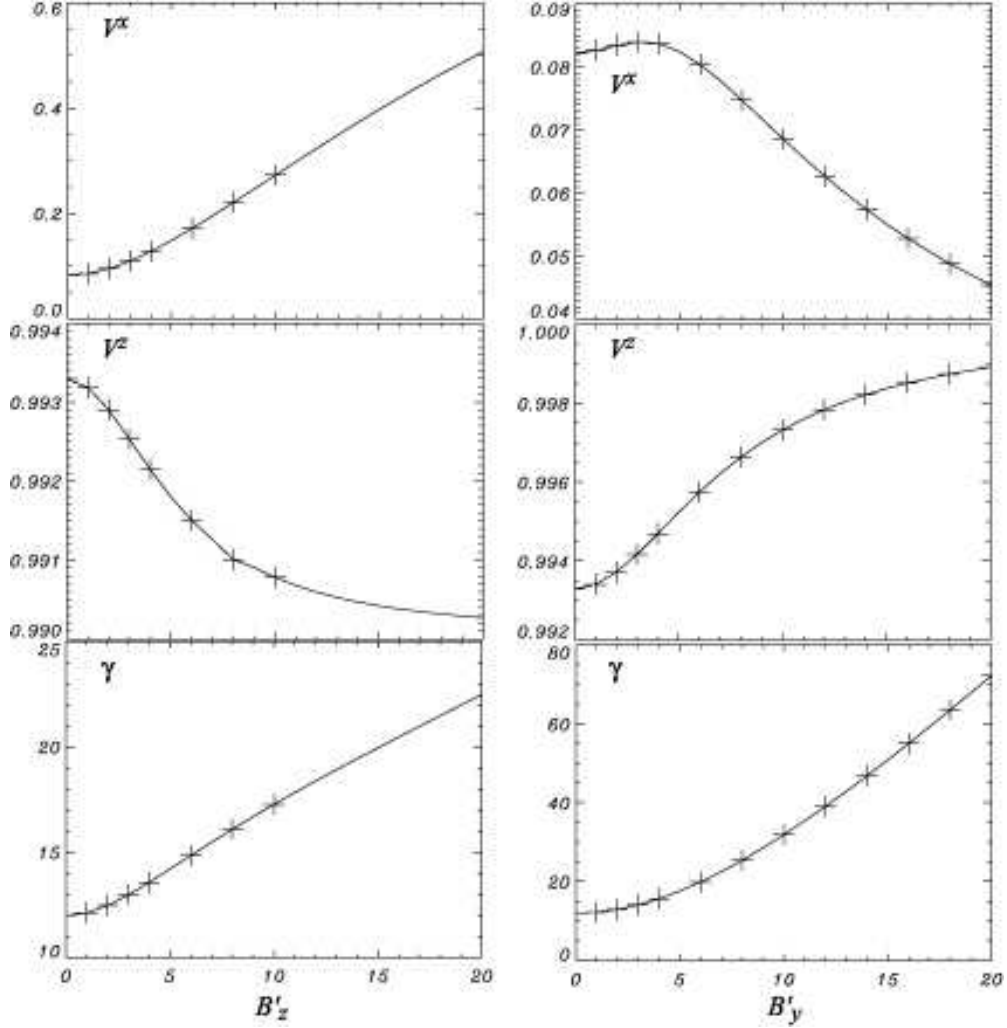}
\caption{Dependence of maximum
normal velocity ($v^{x}$) ({\it upper panel}), maximum
tangential velocity ($v^{z}$) ({\it middle panels}) and maximum
Lorentz factor $\gamma = [1-(v^{x})^{2}-(v^{z})^{2}]^{-1/2}$
({\it lower panels}) on the strength of $z$-component of magnetic
field $B^{'}_{z}$ ({\it left panels}) and $y$-component of magnetic
field $B^{'}_{y}$ ({\it right panels}). The solid line indicates values
obtained using the code of Giacomazzo \& Rezzolla (2006) and
the crosses indicate the values obtained from our simulations
at the time $t=0.2$. \label{f4}}
\end{figure}

\begin{figure}
\epsscale{0.9} 
\plotone{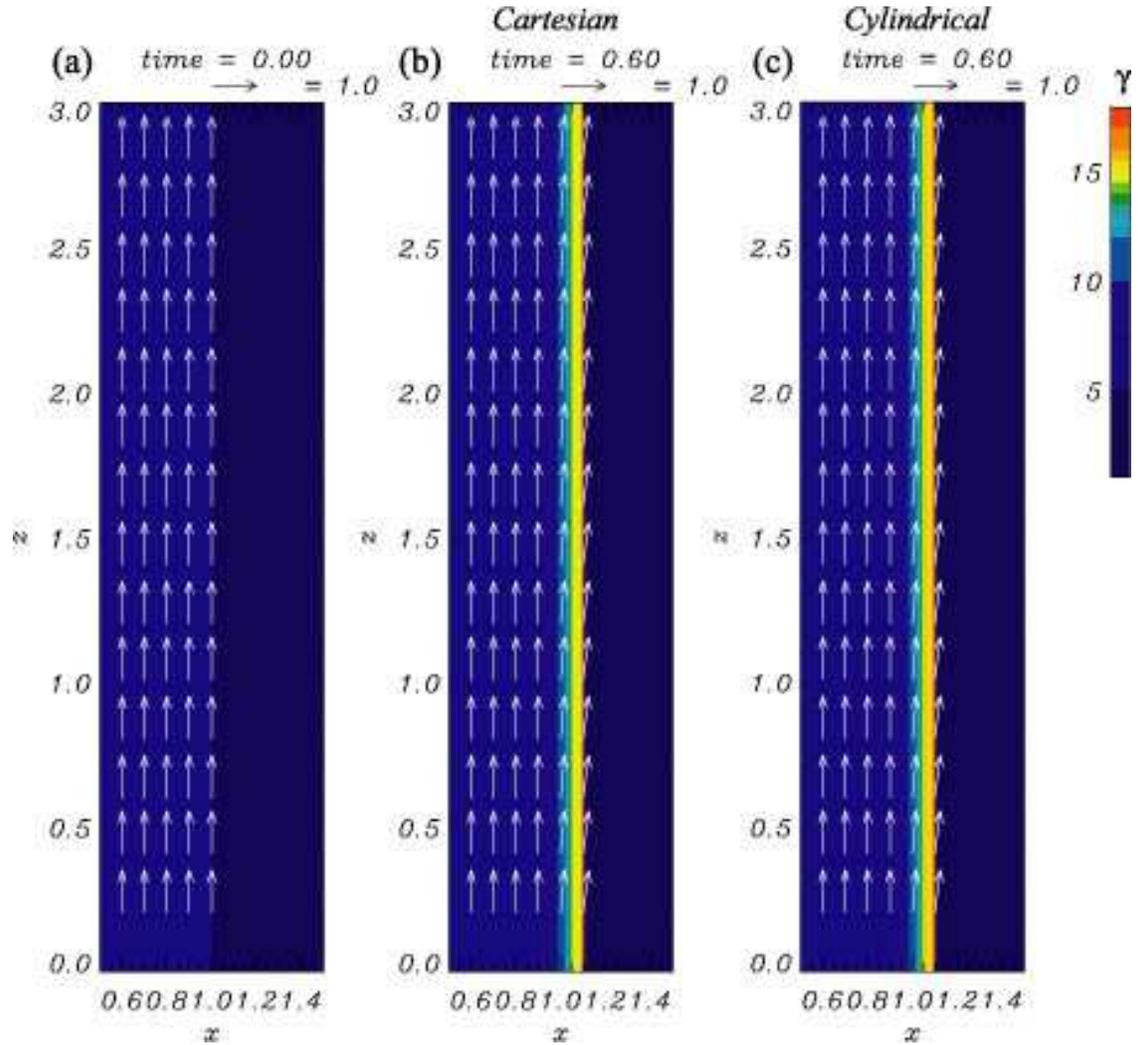} 
\caption{2D images of the Lorentz factor showing
the ({\it a}) initial condition, ({\it b}) results of the 2D MHDA case in
Cartesian coordinates and ({\it c}) results of the
2D MHDA case in cylindrical coordinates at time $t=0.6$. The color
scales show the Lorentz factor. Arrows depict the poloidal
velocities normalized to light speed. \label{f5}}
\end{figure}

\begin{figure}
\epsscale{0.7} 
\plotone{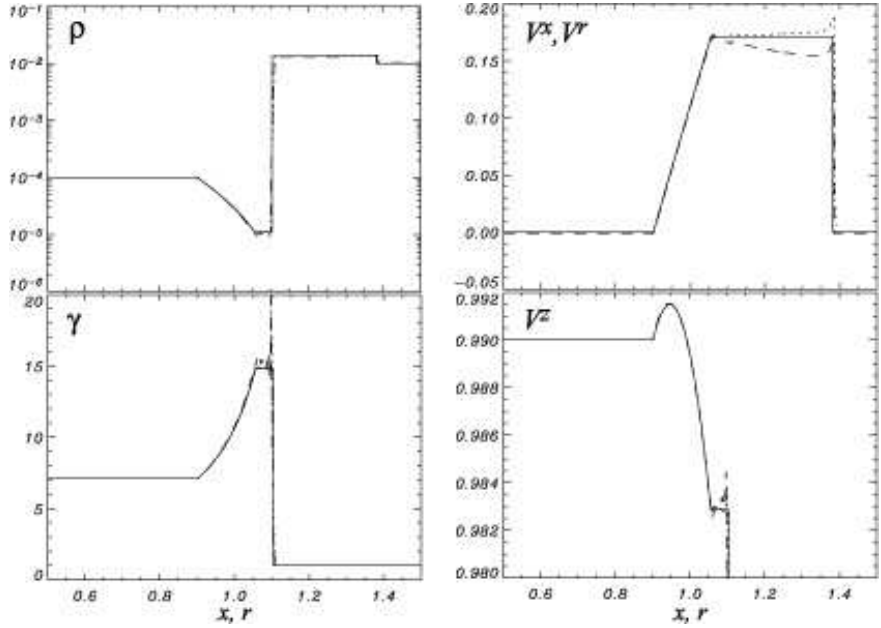} 
\caption{Profiles of ({\it left-upper panel})
density, ({\it left-lower panel}) Lorentz factor, ({\it right-upper panel})
normal velocity ($v^{x}, v^{r}$), and ({\it right-lower panel}) tangential velocity
($v^{z}$) of the 2D MHDA case in Cartesian coordinates
({\it dotted lines}) and the 2D MHDA case in cylindrical coordinates
({\it dashed lines}) at time $t=0.6$. The solid lines are the exact 
solution of the 1D MHDA case at time $t=0.6$. \label{f6}}
\end{figure}

\begin{figure}
\epsscale{0.85}
\plotone{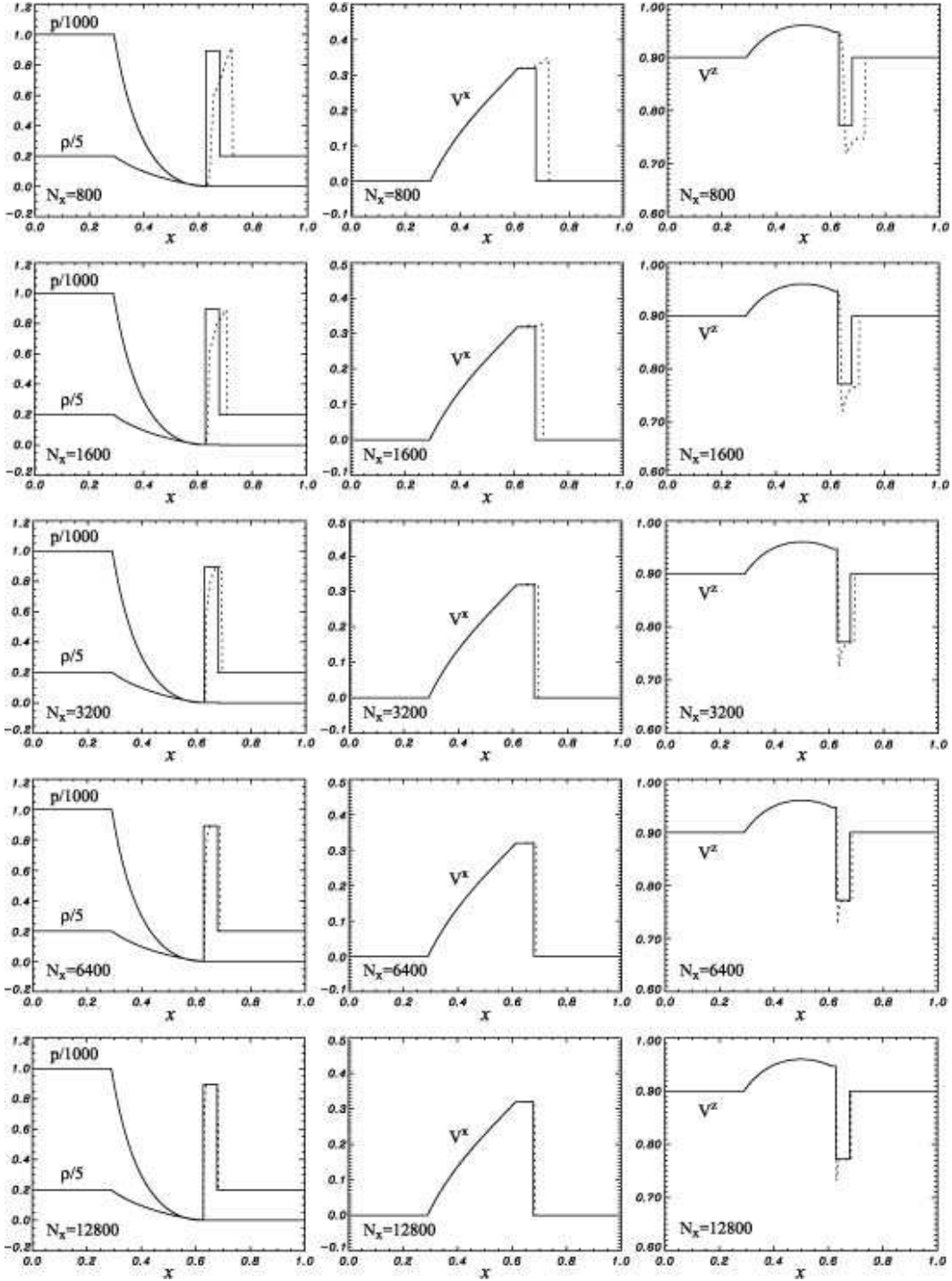} 
\caption{ Profiles of the shock tube test problem with a transverse
velocity at time $t=0.4$. The solid lines are the exact solution and
the dotted lines are the simulation results.  Different simulation
resolutions are presented; the number of computational zones ($N_{x}$)
is 800, 1600, 3200, 6400, and 12,800 from top to bottom.  The density,
gas pressure, normal velocity ($v^{x}$) and tangential velocity
($v^{z}$) are shown. \label{fap1}}
\end{figure}

\begin{figure}
\epsscale{0.85} 
\plotone{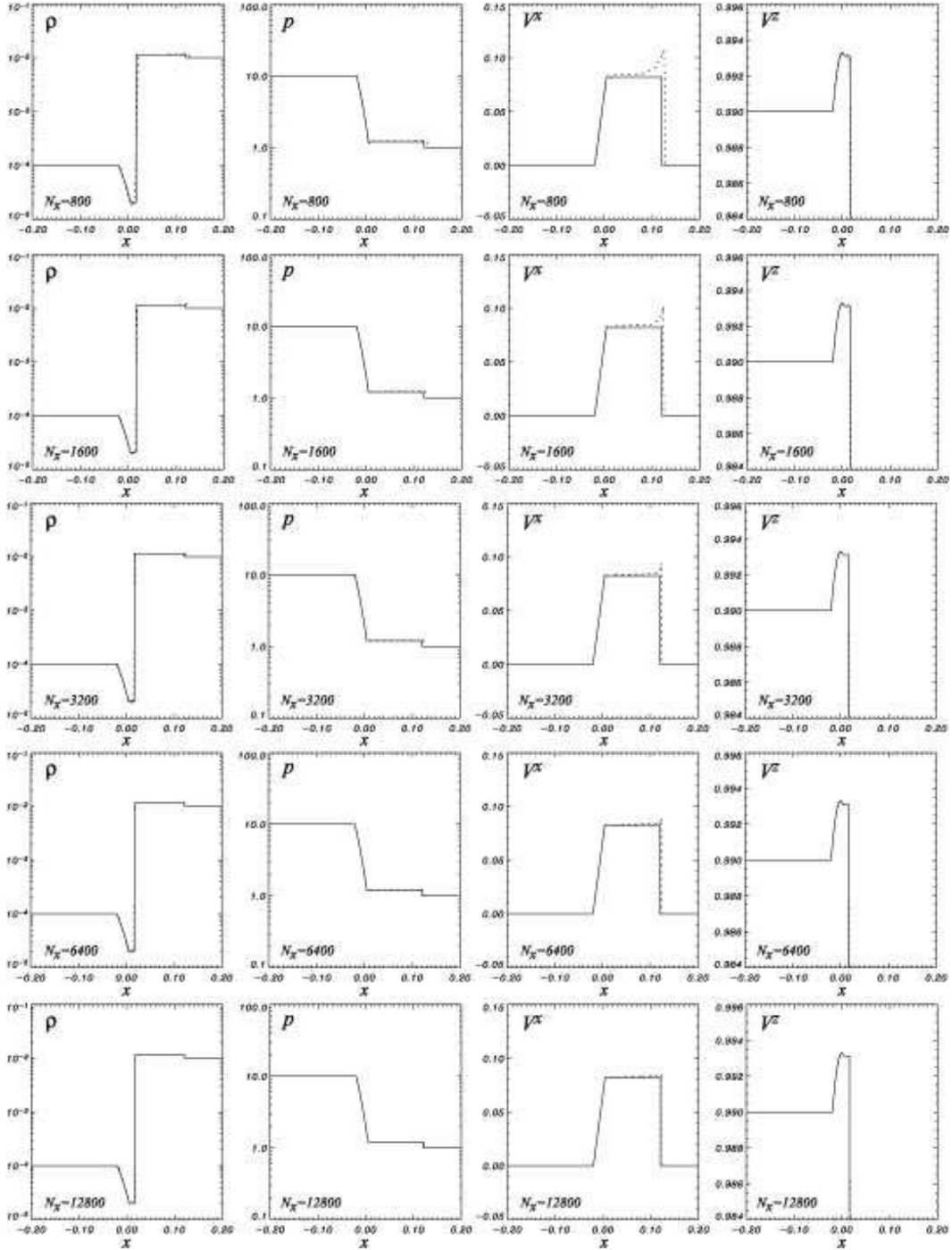} 
\caption{ Profiles of the HDA case at time $t=0.2$. The solid lines
are the exact solution and the dotted lines are the simulation
results.  Different simulation resolutions are presented; the number
of computational zones ($N_{x}$) is 800, 1600, 3200, 6400, and 12,800
from top to bottom.  The density, gas pressure, normal velocity
($v^{x}$) and tangential velocity ($v^{z}$) are shown. \label{fap2}}
\end{figure}

\begin{figure}
\epsscale{0.85} 
\plotone{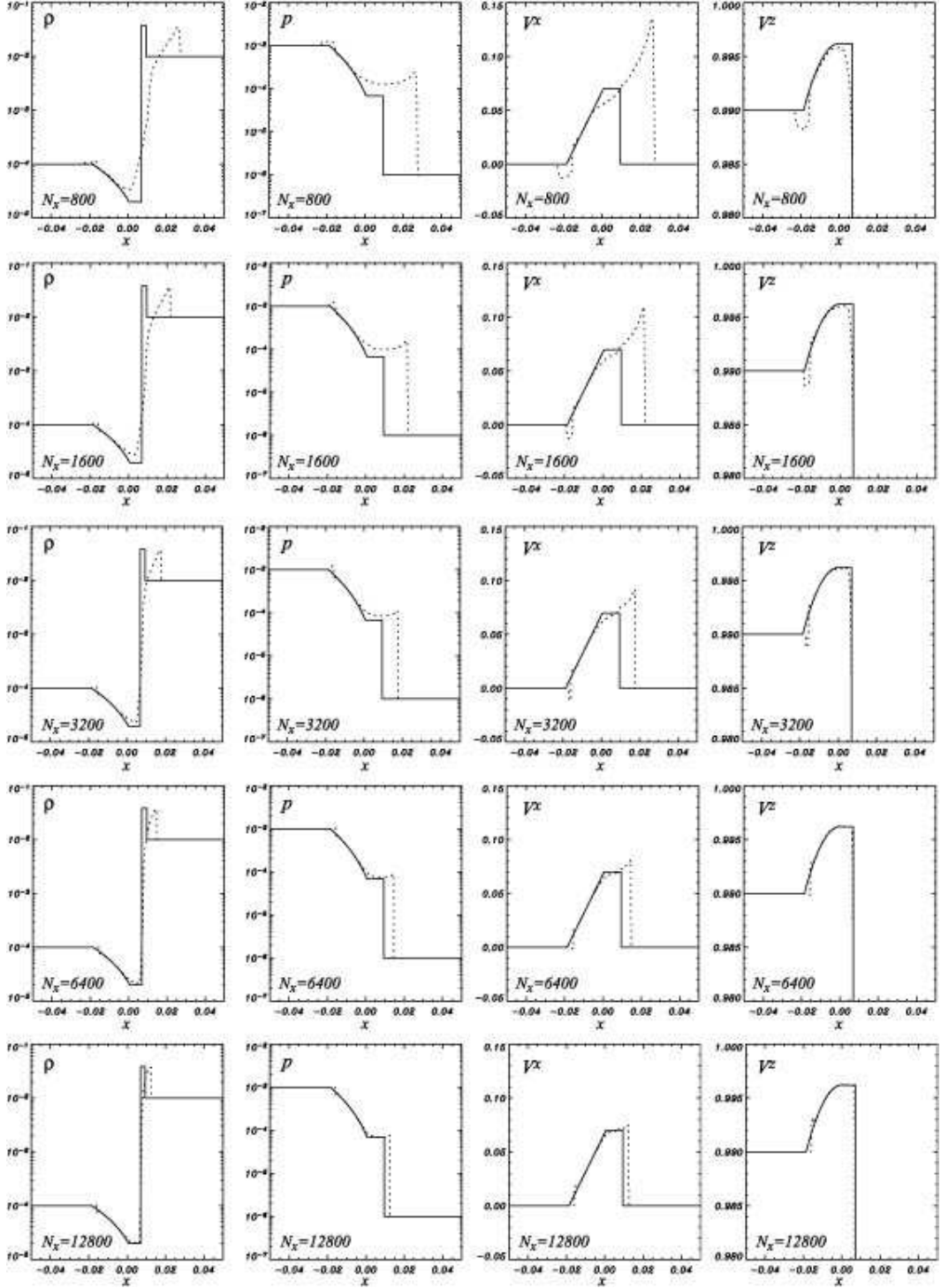} 
\caption{ Profiles of hydrodynamic relativistic boost model proposed by 
Aloy \& Rezzolla (2006) at time $t=0.2$. The solid lines
are the exact solution and the dotted lines are the simulation
results.  Different simulation resolutions are presented; the number
of computational zones ($N_{x}$) is 800, 1600, 3200, 6400, and 12,800
from top to bottom.  The density, gas pressure, normal velocity
($v^{x}$) and tangential velocity ($v^{z}$) are shown. \label{fap3}}
\end{figure}

\end{document}